\documentclass[aps,prb,twocolumn,groupedaddress,floatfix]{revtex4-1}
\usepackage{amssymb,amsmath}
\usepackage{graphicx}
\usepackage{subfigure}
\usepackage[english]{babel}
\usepackage{float}
\usepackage{color}

\begin{document}
\newcommand{\be}{\begin{equation}}
\newcommand{\ee}{\end{equation}}
\newcommand{\rojo}[1]{\textcolor{red}{#1}}

\title{Saturable impurity in an optical array: Green function approach}

\author{Mario I. Molina}
\affiliation{Departamento de F\'{\i}sica, Facultad de Ciencias, Universidad de Chile, Casilla 653, Santiago, Chile}

\date{\today }

\begin{abstract} 

We examine a one-dimensional linear waveguide array containing a single saturable waveguide. By using the formalism of lattice Green functions, we compute in closed form the localized mode and the transmission across the impurity in closed form.
For the single saturable impurity in the bulk, we find that an impurity state is always possible, independent of the impurity strength. For the surface saturable impurity case, a minimum nonlinearity strength is necessary to create a bound state. The transmission coefficient across the impurity shows a sub-linear behavior with an absence of any resonance.  The dynamical selftrapping at the bulk impurity site shows no selftrapping transition, and it resembles the behavior of a weak linear impurity. For the surface impurity however, there is a selftrapping transition at a critical nonlinearity value. The asymptotic propagation of the optical power shows a ballistic character in both cases, with a speed that decreases with an increase in nonlinearity strength.

\end{abstract}

\maketitle

\section{Introduction}
The effect of one or few impurities embedded in a periodic system is an old problem, whose interest has not vaned throughout the years\cite{defect1, defect2}. For a discrete system such as a chain of atoms or an optical  waveguide array, a linear isolated impurity breaks the translational invariance and gives rise to an exponentially decreasing  localized mode centered at the impurity site, no matter how small the impurity strength\cite{TB1, TB2}. Other kind of defects include coupling defects, junction defects between two optical or network arrays\cite{junction}, discrete networks for routing and switching of discrete optical solitons\cite{miro}, and also in simple models for magnetic metamaterials, modeled as periodic arrays of split-ring resonators, where magnetic energy can be trapped at impurity positions\cite{SRR}

When nonlinearity is added to a periodic waveguide array, mode localization and selftrapping of energy can occur. This localized mode which exists in this nonlinear but completely periodic system is known as a discrete soliton. This concentration of energy on a small region increases with the nonlinearity strength and, as a consequence the nonlinear mode becomes effectively decoupled from the rest of the lattice. This is very similar to the case where one has a cluster of few impurities sites (or even a single one) embedded in an otherwise linear chain. In this case, the problem becomes much simpler and, sometimes, closed form results can be obtained. 

A common method for dealing with impurity problems is to make an educated guess about the impurity profile. This procedure usually works fine with linear impurities, but when nonlinearities enter the picture, it is no longer certain that this method will work in all cases. One elegant method for dealing with impurity problems is the technique of lattice Green functions\cite{economou,barton,duffy}. Even though this formalism was originally derived for linear problems, it has been shown that it can also be extended to simple nonlinear problems\cite{molina_green1,molina_green2,molina_green3,molina_green4}.

In this work we consider a single saturable impurity inside the bulk and at the surface of a one-dimensional linear waveguide array. By using an  extension of the usual formalism of lattice Green functions, we compute in closed form the energy of the localized mode, its spatial profile, and the transmission coefficient of plane waves across the saturable impurity. For the impurity in the bulk we find that there is no minimum nonlinearity strength to effect a bound state while for the surface impurity the nonlinearity strength must exceed a critical amount to create a surface bound state. 

\section{The model}
\begin{figure}[t]
 \includegraphics[
 scale=0.25]{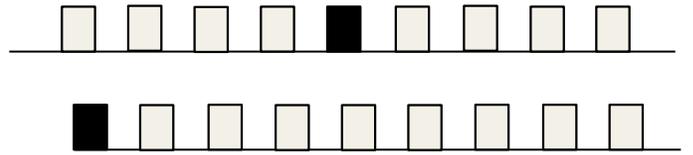}
  \caption{Schematic view of an optical waveguide array containing a saturable impurity in the bulk (top) and at the surface (bottom) of the array.}
  \label{fig1:dimm}
\end{figure}

Let us consider an weakly-coupled, optical waveguide array containing a single saturable impurity waveguide in  the bulk and at the surface of the array (Fig.1). While the linear guides are usually made from semiconductor materials ($\mbox{GaAs/AlGaAs}$), the saturable waveguide can be fabricated from  lithium niobate doped with iron ($\mbox{Fe:LiNbO}_{3}$), for instance. In the coupled-mode formalism, the dimensionless equations for the evolution of the electric field amplitude at the $n$th guide are
\be 
i {d E_{n}\over{d z}} + V (E_{n+1} + E_{n-1}) + \delta_{n,d} {\chi \ E_{n}\over{1 + |E_{n}|^2}},\label{1}
\ee
with $n=1,2,\cdots,N$, and $d$ is the position of the impurity guide: $d=0$ for the surface impurity, and $d \approx (N/2)$ for the bulk case.  Parameter $z$ is the dimensionless distance along the longuitudinal direction and $V$ is the coupling parameter. The presence of the impurity tends to change the coupling between the impurity and its nearest-neighbor guides. This can be compensated by an adequate shifting of the distances between the impurity guide and its nearest neighbors, to ensure the same coupling $V$ for all guides.

\subsection{Bulk impurity}

In the Green function approach, we start from the Hamiltonian of the system, which in our case can be written as
\be
\tilde{H} = \tilde{H_{0}} + \tilde{H_{1}}
\ee
\be
\tilde{H_{0}}=V \sum_{nn}(|n\rangle \langle m| + h.c.)
\ee
\be
\tilde{H_{1}} = {\chi\over{1+|E_{d}|^2}} |d\rangle \langle d|
\ee
where $E_{d}$ is the amplitude at the impurity, which has been placed at $n=d$ with $0\ll d \ll N$ and we use the Dirac notation for convenience. Here, the $\{|n\rangle \}$ represent Wannier-like states. Now we normalize all energies to a half bandwidth, $4 V$ and define $z\equiv E/4 V$, $H\equiv \tilde{H}/4 V$, and $\gamma\equiv \chi/4 V$. Now the system Green function, $G=1/(z-H)$ can be expanded as
\be
G=G^{(0)} + G^{(0)} H_{1} G^{(0)} + G^{(0)} H_{1} G^{(0)} H_{1} G^{(0)} + \cdots
\ee
where $G^{(0)}$ denotes the unperturbed Green function $G^{(0)}=1/(z-H_{0})$ and $H_{1}=\gamma/(1+|E_{d}|^2)$. The perturbative series can be summed to all orders to give
\be
G_{m n}=G_{m n}^{(0)} + {\varepsilon\over{1 - \varepsilon\ G_{d d}^{(0)}}} \ G_{m d}^{(0)} G_{d n}^{(0)},
\ee
where $G_{m n} = \langle m |G|n\rangle$ and $\varepsilon=\gamma/(1+|E_{d}|^2)$. Typically, the presence of an impurity in a periodic system, gives rise to a localized mode around the impurity, whose energy lies outside of the band. As we will see, our case is not the exception, and we will denote the energy of this state as $z_{b}$ and its amplitudes as $E_{n}^{(b)}$.
It should be 
remarked that we don't know $G_{nm}$ yet 
because we don't know $|E_{d}|^2$. We shall obtain it by the self-consistent procedure described below.

The bound state energy $z_{b}$ is given by the poles of $G_{m n}$, i.e., by solving 
\be
1=\varepsilon\  G_{d d}^{(0)}(z_{b})= \gamma\ {G_{d d}^{(0)}(z_{b})\over{1+|E_{d}^{(b)}|^2}}\label{7}
\ee
where the unperturbed Green function is $G_{n d}^{(0)}(z) = (\mbox{sgn(z)}/\sqrt{z^2-1}) \{z - \mbox{sgn(z)} \sqrt{z^2-1}\}^{|n-d|}$ for $z$ outside the band.
On the other hand, the bound state amplitudes $E_{n}^{(b)}$ are given by the residues of $G_{m n}(z)$ at $z=z_{b}$
\be
|E_{n}^{(b)}|^2 = \mbox{Res} \{ G_{n d}(z) \}_{z=z_{b}} = -{G_{n d}^{2}{(z_{b})}\over{G'_{d d}(z_{b})}}.\label{e02}
\ee
\begin{figure}[t]
\includegraphics[scale=0.45]{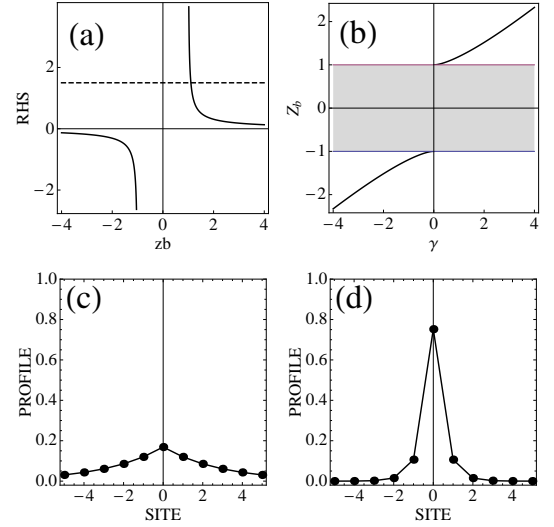}
\caption{(a) RHS of Eq.(9) as a function of $z$. The intersection with the line $1/\gamma$ determine the localized mode energy. (b) Bound state energy as a function of nonlinearity. (c) and (d): Spatial profiles for a mode with $\gamma=0.2$ (left) and $\gamma=2$ (right). We have taken $d=0$.}
\label{fig2}
\end{figure}
Inserting this into Eq.(\ref{7}) (after setting $n=d$) leads to an  equation for the bound state energy:
\be
{1\over{\gamma}} = {{G_{d d}^{(0)}}(z) \ G_{d d }'^{(0)}(z)\over{G_{d d}'(z)- G_{d d}^{2}(z)}}\label{9}
\ee
that is,
\be
{1\over{\gamma}} = {z\over{z^2 -1 + |z|\sqrt{z^2-1}}}.\label{10}
\ee
Figure 2a shows the plot of $1/\gamma$ and the RHS of Eq.(\ref{10}). Clearly, there is a single real solution for any $\gamma$. Equation (\ref{10}) is a cubic equation, with solution
\be
z_{b} = -\left( {1-\gamma^2\over{6 \gamma}}\right) + {1+10 \gamma^2 + \gamma^4\over{6 \gamma D(\gamma)}}+ {D(\gamma)\over{6 \gamma}}\label{zb}
\ee
where, 
\be
D(\gamma)=-1+39 \gamma^2+15 \gamma^4+\gamma^6+6 \sqrt{3}  \gamma \sqrt{-1+11 \gamma^2+\gamma^4}.
\ee
Figure 2b shows the bound state energy as a function of nonlinearity and we can see that it always lies outside the linear band. In Fig. 2c, 2d we show the spatial profiles for two different nonlinearity strengths. These profiles are given in closed form by
\be
|E_{n}^{(b)}|^2 = {\sqrt{z_{b}^2-1}\over{|z_{b}|}} \left( z_{b}-\mbox{sgn}(z_{b})\sqrt{z_{b}^2-1} \right)^{2 |n-d|}
\ee
where $z_{b}$ is given by Eq.(\ref{zb}). The spatial decay is exponential $\sim \exp(-|n-d|/\lambda)$, with a localization length $\lambda$ given by $\lambda=(2 \log(1/\alpha))^{-1}$, where $\alpha=z_{b}-\mbox{sgn}(z_{b})\sqrt{z_{b}^2-1}$.

\subsection{Surface impurity}
In this case, the impurity is located at one of the boundaries of the 1D waveguide array, say $d=0$. The only difference with the previous case is that the unperturbed Green function $G_{m n}^{(0)}$ for the semi-infinite array must now take into account the presence of the boundary. This can be done with the method of mirror images. The absence of any waveguide to the left of $n=0$, means that $G_{m n}^{(0)}$ should vanish at $n=-1$. This means, 
$G_{m n}^{(0)}=G_{m n}^{\infty}-G_{m,-n-2}^{\infty}$, where $G_{m n}^{\infty}$ is the unperturbed Green function for the infinite system we used in the previous section. Therefore,
\begin{eqnarray}
G_{m n}^{(0)}& = &{\mbox{sgn(z)}\over{\sqrt{z^2-1}}}\left[ z -\mbox{sgn(z)}\sqrt{z^2-1}\right]^{|n-m|}\nonumber\\
& & -{\mbox{sgn(z)}\over{\sqrt{z^2-1}}}\left[ z -\mbox{sgn(z)}\sqrt{z^2-1}\right]^{|n+2+m|}.
\end{eqnarray}
Using this unperturbed Green function the eigenvalue equation (\ref{9}) becomes
\be
{1\over{\gamma}}={2\over{z+3\ \mbox{sgn(z)} \sqrt{z^2-1}}}\label{16}
\ee
A simple graphical analysis (Fig. 3a) shows that a minimum nonlinearity $|\gamma|=1/2$ is needed to create a bound state. Figure 3b shows the bound state energy as a function of nonlinearity. It is given by $z_{b}=(1/4)(-\gamma+3\ \mbox{sgn}(\gamma) \sqrt{2+\gamma^2})$. 
From Eq.(\ref{e02}) the bound state spatial profile is given by
\be
|E_{n}^{(b)}|^2 = \alpha(z_{b}) (\  q(z_{b})^{|n|} - q (z_{b})^{|n+2|} \ )
\ee
where $n=0,1,2,\cdots$ and 
\be
q(z)\equiv z - \sqrt{z^2-1}, \hspace{0.2cm}\alpha(z)\equiv 2 - 2\ z\ (z-\sqrt{z^2-1}).
\ee
This profile is no longer a simple exponential decay, but is the superposition of two such decays. Figures 3c, 3d show a couple of examples of these profiles.
\begin{figure}[t]
\includegraphics[scale=0.45]{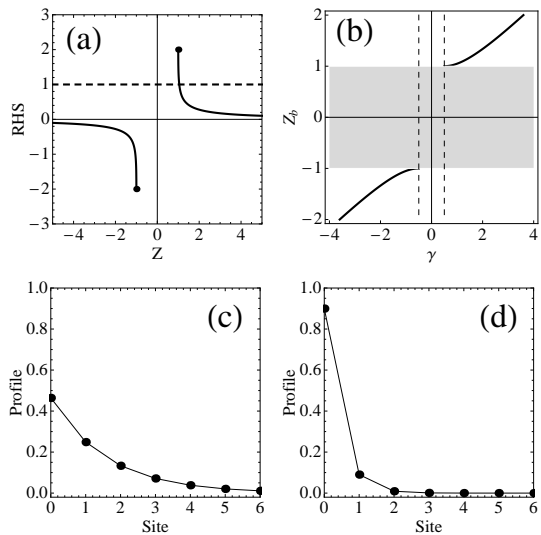}
\caption{(a) RHS of Eq.(\ref{16}) as a function of $z$. The intersection with the line $1/\gamma$ determine the localized mode energy. (b) Localized mode energy as a function of nonlinearity. (c) and (d): Spatial profiles for a mode with $\gamma=1$ (c) and $\gamma=3$ (d)}
\label{fig3}
\end{figure}

\subsection{Transmission}
Let us now consider the scattering of planes across the saturable impurity.
The scattering states inside the band are given by \cite{economou}
\be
\langle n|E\rangle = \langle n|k\rangle+ \varepsilon\  {\langle n|G^{(0)+}(z)|d\rangle \langle0|k\rangle\over{1-\varepsilon\  \ G_{d d}^{(d)}}}
\ee
where $\varepsilon=\gamma/(1+|E_{d}|^2)$. The first term is the incoming wave and the second one is the scattered wave.
The amplitude at the impurity guide is given by,
\be
E_{d}=1 + {\varepsilon\ G_{d d}\over{1-\varepsilon\ G_{d d}}}={1\over{1-\varepsilon\ G_{d d}}}
\ee
The transmission coefficient $t$ is the square of the amplitude at the impurity guide, $t=|E_{d}|^2$, that is,
\be
t = |1 - \varepsilon\ G_{d d}^{(0)}|^{-2} = \left|1 - {\gamma\over{1 + t}}\ G_{d d}^{(0)}\right|^{-2}\label{21}
\ee 
Using $G_{d d}^{(0)}=i/ \sqrt{1-z^2}$, we get a quadratic equation for t, with real solution
\be
t={1\over{2}}\left( -\gamma^2+\sqrt{4 + \gamma^4-8 \cos^{2}(k)+ 4 \cos^{4}(k)}\ \right) \csc^{2}(k)
\ee

Figure 4a shows the transmission coefficient as a function of wavevector, for several different nonlinearity strengths. Since the ``effective'' impurity parameter $\gamma/(1+t)$ is always smaller that $\gamma$, the saturable impurity is always ``weaker'' than a linear one. Therefore, the saturable transmission is always larger than its linear counterpart.

\subsection{Dynamic properties}
Let us first consider the possible seltrapping of optical power at the (bulk and surface) saturable guide. We place all power at the impurity at $t=0$ and observe the time evolution of $|E_{d}(z)|^2$ for long times. In order to avoid undesirable reflections from the boundaries, we used a self-expanding lattice. To see the presence of selftrapping, we compute the long-time average power at the impurity guide, defined as
\be
P_{d} = \lim_{T\rightarrow \infty} (1/L) \int_{0}^{L} |E_{d}(z)|^2 dz.
\ee
\begin{figure}[tbh]
\includegraphics[scale=0.45]{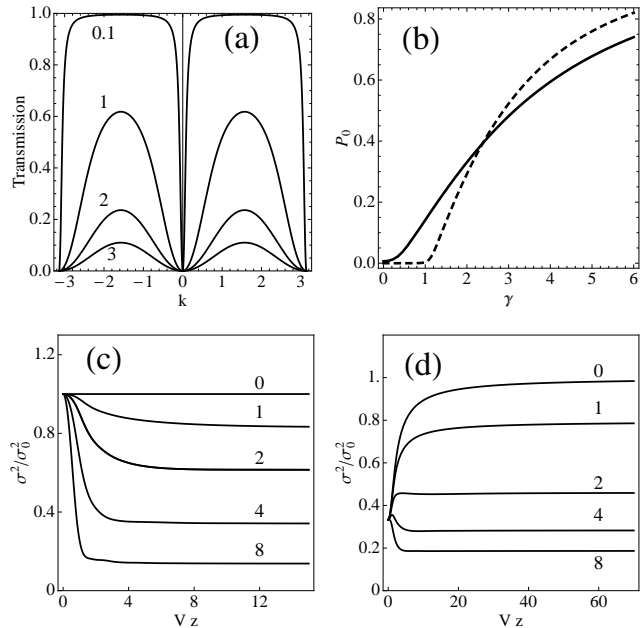}
\caption{(a) Transmission  across the impurity for several nonlinearity values (b) Trapping of optical power at the impurity guide as a function of nonlinearity. The continuous (dashed) curve refer to the bulk (surface) case. (c) and (d): Evolution of the mean square displacement for the bulk (c) and surface (d) cases.}
\label{fig4}
\end{figure}
Figure 4b shows the average trapped fraction of optical power on the impurity guide, for both, bulk and surface cases. For the bulk case we observe a smooth increase of $P_{d}$ with $\gamma$. For the surface case however, $P_{d}$ is esentially zero, until around $\sim 1$, there is a sudden increase in $P_{d}$, that is, there is a seltrapping transition.

Let us now examine the lateral propagation of the optical power, quantified by the mean square displacement,
\be
\langle n^2 \rangle = {\sum_{n}{(n-d)^2 |E_{n}(z)|^2}\over{\sum_{n} |E_{n}(z)|^2}}
\ee
We know that, for a completely localized initial condition, and in the absence of the impurity guide, the lateral propagation is ballistic, i.e., $\langle n^2 \rangle = 2 (V z)^2$ for an initial excitation in the bulk of the (wide) array, or $3 (V z)^2$ when the initial excitation is placed at the boundary of the array\cite{Martinez}. Figures 4c and 4d show $\langle n^2 \rangle$ as a function of evolution distance $z$, for the bulk and surface saturable impurity. In both cases we observe that, after a transient evolution, both cases revert to the ballistic case, although with ``speeds'' that decrease with an increase of the impurity nonlinearity strength. This can be easily understood as a consequence of the partial seltrapping at the initial site which renormalizes  the total optical power that can propagate to infinity. 

\subsection{Conclusions}
By using the formalism of lattice Green functions we have obtained in closed form the nonlinear modes and 
the  transmission coefficient across a saturable impurity emdedded in a linear waveguide array. When the impurity guide is placed at the array bulk there is a bound state for any nonlinearity strength while for the surface case there is a minimum strength required. The transmission across the bulk impurity shows no resonances and it resembles a linear transmission. The selftrapping at the initial site shows no transition for the bulk case, but there is minimum nonlinearity strength to effect selftrapping for the surface impurity. Finally, the long-time propagation of optical power shows a ballistic behavior, with a speed that decreases with an increase in nonlinearity. All in all, the behavior of this saturable impurity is reminiscent of the case of a linear impurity, due to tha fact that its effective nonlinearity is always smaller than the corresponding linear counterpart.

\acknowledgments
This work was supported by Fondecyt Grant 1160177.

\end{document}